\documentclass{appolb}
\usepackage{epsfig}

\newcommand{\ra}{{\rightarrow}}
\newcommand{\vev}[1]{\langle #1 \rangle}

\begin{document}
\title{Particle Production and Propagation in Nuclei\thanks{Presented at the
International Conference on Multiparticle Production ISDM 2003 in Cracow}}
\author{H.J.~Pirner
\address{Institut f\"ur Theoretische Physik der Universit\"at Heidelberg\\
Philosophenweg 19, D-69120 Heidelberg, Germany\\
E-mail address: pir@tphys.uni-heidelberg.de\\
and\\
Max-Planck-Institut f\"ur Kernphysik, Heidelberg, Germany}}
\maketitle
\begin{abstract}
We discuss the effects of gluon radiation by the struck quark
and the subsequent absorption of the produced hadron in deep 
inelastic lepton-nucleus scattering. 
The theoretical picture is compared with HERMES results on
multiplicity ratios. 
\end{abstract}
  
\vspace{1.0cm}

The fragmentation of light quarks  is still not completely understood.
The  nucleus helps 
to understand the space time evolution of a parton, since the nucleons
play the role of very nearby detectors of the propagating object. 
Deep inelastic electron nucleus scattering has the advantage that the
electron gives a well defined energy $\nu$ to the struck quark 
propagating through cold nuclear matter. The understanding of this
process is crucial for the interpretation  of  
ultra-relativistic proton-nucleus and nucleus-nucleus collisions.
Due to factorization in deep inelastic scattering the semi-inclusive
cross section can be described by the product of a parton distribution function
(PDF)
with  a fragmentation function (FF) cf. Fig.~1.  
In the parton model the probability  
that a  quark
with momentum fraction $x$  
is present in the target is multiplied with  the probability 
that it hadronizes 
into a definite hadron which carries a momentum fraction $z$ of 
the quark.
Inclusive electron proton and electron nucleus scattering allows to
measure the structure function in both types of targets. 
Quark hadronization and hadron production 
can be studied independently in $e^+e^-$ annihilation.
Fig.1 shows a schematic diagram of semi-inclusive
deep inelastic lepton scattering (SIDIS) on a target $T$, and the definitions
of the four momenta of the particles involved in the process.
In SIDIS besides the scattered lepton $l\,'$
the leading hadron $h$ formed from the struck quark is detected 
with energy $E_h=z\nu$ in the target rest frame.
The summation over flavors includes the product of 
the fragmentation functions and structure functions for each flavor.
The experimental data on nuclear effects in hadron production 
are usually presented in terms of multiplicity ratios  as 
functions of $z$,$\nu$ or recently $Q²$:
\begin{equation} 
    R_M^h(z)  = 
        \frac{1}{N_A^\ell} \frac{dN_A^h}{dz}
        \bigg/ \frac{1}{N_D^\ell} \frac{dN_D^h}{dz}. 
 \label{multrationu}
\end{equation}
In the above definitions $N_A^\ell$ is the number of outgoing leptons
in DIS processes on 
a nuclear target of atomic number $A$, while $dN_A^h/dz$ is
the $z$- distribution of produced hadrons in the same
processes; the 
subscript $D$ refers to the same quantities when the target is a
deuteron.  
In absence of nuclear effects the ratio $R^h_M$ would be equal to
1. 

\begin{figure}[t]
\begin{center}
\parbox{7.5cm}{\vskip-.5cm
\epsfig{figure=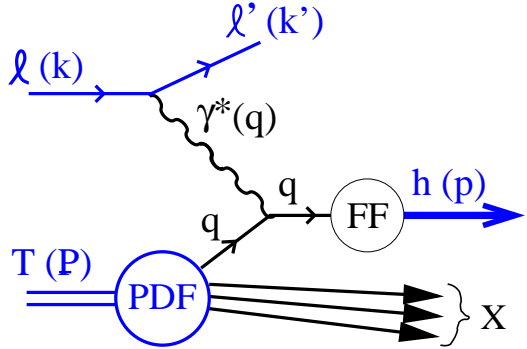,width=7cm}}
%
\vskip0cm
\caption{\footnotesize 
Semi-inclusive hadron production in deep inelastic
scattering on a target T in the pQCD factorization approach. Parton 
distribution functions (PDF) and fragmentation
functions (FF) represent the non-perturbative
input.}
\label{fig:DIS}
\end{center}
\vskip-.4cm
\end{figure}
                     
The HERMES \cite{Hermes} experiment has brought new insight into the question, since
it has access to
lower energy transfers $\nu$, where the hadronization
occurs inside the nucleus,and to a larger range of fractional momenta $z$
which determine the formation length of the hadron. 
In this paper we give an introductory overview 
of the different models used in describing 
hadron production in deep inelastic scattering.
These include ``QCD inspired'' analyses such as
the rescaling model \cite {Accardi:2002tv}, 
energy loss models (with and without higher twist
\cite{Baier:1996sk,Arleo:2003jz,hightwist}),
the gluon-brems\-strah\-lung calculation 
for leading hadron production \cite{KNP2003},
and  nuclear absorption of the prehadron 
\cite{Accardi:2002tv,BC83}. 
For a survey of other approaches we refer to \cite{Muccifora02}.
We also would like to mention the recent review by the CERN study
group on ``Hard Probes in Heavy Ion Collisions at the LHC: Jet
Physics''\cite{Accardi:2003gp},
which covers the problem of hadronization in a  hot medium.
This short note will only address the problem of hadronization
in deep inelastic scattering.

Parton distribution functions and fragmentation functions
 both depend on the virtuality $Q^2$ of the DIS process.
This adjustment to  the scale $Q^2$ takes into account all 
radiated gluons before and after the photon quark interaction in 
the leading logarithm approximation. In nuclei gluon radiation may be affected
by the partial deconfinement of color which follows from overlapping
nucleons. The long wavelength spectrum  of gluons extends farther into
the infrared towards low $Q^2 \propto 1/\lambda^2$ where $\lambda$ is
the confinement scale.

Therefore in DGLAP evolution in 
nuclear structure and fragmentation functions the starting scale is smaller
and they evolve over a larger interval in momentum compared with 
the corresponding
functions in the nucleon at the same scale Q. The solution of 
the DGLAP equation gives  $q^A_f(x,Q) = q_f(x,\xi_A(Q) Q)$
and $D^{h|A}_f(z,Q) = D^h_f(z,\xi_A(Q) Q)$, where the scale 
factor $\xi_A(Q)$ is related to
the deconfinement scale
$\lambda_A$ which is proportional to the
overlap of nucleons inside the given nucleus. 
Typical values of the scale factor are $\xi_A=1.2$ for Kr or Xe. 
The multiplicity ratio is calculated by using the rescaled 
parton distribution functions 
and the rescaled fragmentation functions.
Because of the $\nu$ dependence of the 
fragmentation process in DIS on nuclei the assumption of factorization
is certainly open to criticism. Most models do not derive
the fragmentation process from scratch and only manipulate the
factorized  formula
\begin{equation}
    \frac{1}{N_A^\ell}\frac{dN_A^h}{dz} = \frac{1}{\sigma^{\gamma^*A}} 
         \int_{\rm exp.\ cuts}\hspace*{-1.0cm}
         dx\, d\nu \sum_f e^2_f\, q_f^{(A,Z)}(x,\xi_A\,Q) \,
         \frac{d\sigma^{\gamma^* q}}{dx\,d\nu} \, D_f^h(z,\xi_A\,Q).  
\end{equation}
In the formula above $e_f$ is the electric charge of a quark of flavour $f$,  
$d\sigma^{\gamma^* q}/dx\,d\nu$ is the differential cross-section for
a $\gamma^* q$ scattering computed in pQCD at leading order.
Due to Lorentz dilatation, at large values of
$\nu$ the hadrons are expected to form mainly outside the nucleus, so
that the effect of reinteractions of the hadron with the nucleus are
minimal. 
Gluon radiation is also induced by scatterings of the 
struck quark with the nucleons in the nucleus. Various ideas have
been proposed to calculate this process. Medium induced gluon
radiation \cite {Baier:1996sk} leads to an energy loss per unit length
\begin{equation}
       \frac {d E}{dL} =- \frac{3}{2} \alpha_s <p_t^2>_L= - 
\frac{3}{2} \alpha_s
       <p_t^2>_0 \frac{L}{\lambda}
\end{equation}
which is proportional to the acquired $<p_t^2>_L$ along the path
length $L$. Because of the random walk in transverse space the
total transverse momentum can be calculated from each individual
momentum transfer times the number of collisions, which
is related to the mean free path $\lambda$. Numerically in cold matter
the energy loss proposed by various authors 
varies between $0.19 \frac{GeV}{fm}<\frac{d E}{dL} <0.5
\frac{GeV}{fm}$ for $L=5 fm$ cf. ref. \cite {Accardi:2003gp}.
In specific models  there is either a $1/Q^2$ dependence, i.e. a higher
twist dependence, or a logarithmic dependence on $Q^2$ for the energy
loss. If there 
would be only energy loss (which we do not believe) 
and no absorption in the Hermes regime, then typical
numbers on the upper edge of this band are needed
to reproduce the experimental data on $R_M$.
Once a prehadron has been produced, it can be absorbed in the nucleus, 
i.e. make an inelastic collision
and change its z fraction. In the gluon bremsstrahlung model
\cite{KNP2003}
 the 
time scale for prehadron formation is related to the 
gluon coherence time which is determined by the light cone energy
difference between the  gluon quark system and the quark:
\begin{equation}
       t_{coh}= \frac{2 \nu z (1-z)}{k_t^2}
\end{equation}
Therefore the prehadron formation
time is short  for large $z$ and small $z$. 
It increases with z because of Lorentz time dilatation and
it is small for large $z$, since the hadron has to be formed
instantaneously, otherwise the energy loss downstream is too large.
In ref. \cite{KNP2003} the gluon radiation model is constructed
based on this formation time. It has the  nice feature that it is mainly
perturbative
and therefore is  well suited to the large $Q^2$ physics of deep
inelastic scattering. In practice, however, the problem lies in
combining a probability for quark gluon formation which is integrated over
the characteristic resolution of the gluons with the overlap matrix
element between the quark gluon state and the hadron. 
Here further development is needed.

Let me finally come to report about the status of the work 
by A. Accardi, V. Muccifora and myself \cite{Accardi:2002tv}.
In order to treat the 
multiplicity ratios in the low-$\nu$ region adequately, we consider 
the 
formation of the prehadron and its subsequent interaction in 
the nuclear medium.
As the hadron formation length $l_f$ decreases by decreasing $\nu$,
the effect of
nuclear interaction becomes more relevant in the kinematic region of 
HERMES  relative to EMC , and the effect is amplified
in a heavy target as the formation length $l_f$ 
is of the same size as the nuclear radius.
Our theoretical model follows closely
the Lund model for the fragmentation
process \cite{LundBook}.
The space-time development of the fragmentation process 
begins when the quark $q$ is
ejected from a nucleon. 
The quark propagates and the  colour string between the quark and the remnant
breaks into  smaller pieces. 
Hadrons are ordered according to their rank
$i$. Note that the first-rank hadron is always
created at the end after the original quark has traversed a
distance $L$.
\begin{equation}
    L=\frac{\nu}{\kappa}
\end{equation}
where $\kappa$ is the string tension. This length can be very large
but this does not mean that long strings exist in the nucleus,
because the string breaks and prehadrons are formed.
In the original paper \cite{Accardi:2002tv} we choose a prehadron cross section equal to the
hadron cross section and adjusted the string tension to fit the
data. We think it is more realistic  that the prehadron coming
from gluon radiation of the struck quark is smaller than the
final hadron. In a new paper in preparation good results are obtained 
with $\sigma (prehadron)\approx \frac{1}{3}\sigma (hadron)$ and 
keeping the string tension
$\kappa=1 GeV/fm$.
There are two relevant lengths for the
fragmentation process,
the position $l_*$ at which the prehadron 
is formed and
the distance $l_h\leq L $ at which the hadron
is formed.
In the Lund model these two lengths are related:
\cite{LundBook}:
\begin{equation}
    l_h = l_* + z L \ .
  \label{l*vslh}
\end{equation}
At fixed $z$ they both increase linearly with the
virtual photon energy $\nu$. However, as functions of $z$ they 
behave rather differently, especially at $z \ra 1 $, where $l_*\ra 0$ 
and $l_h\ra L$. 
We have calculated 
the prehadronic
formation length $\vev{l_F}$ in the Lund model which behaves as
$\vev{l_F} \ra \frac{1}{\kappa} (1-z)\nu$ as $z\ra1$,  
in agreement with the formation length
suggested by the gluon bremsstrahlung model \cite{KNP2003}.
We introduce
a nuclear absorption factor ${\cal N}_A(z,\nu)$ which
is calculated in the Bialas-Chmaj (BC) model.
It  represents the probability that neither the prehadron 
nor the hadron
has interacted with a nucleon.
The multiplicity ratio then reads
as follows:
\begin{equation}
    \frac{1}{N_A^\ell}\frac{dN_A^h}{dz} = \frac{1}{\sigma^{\gamma^*A}} 
         \int_{\rm exp.\ cuts}\hspace*{-1.0cm}
         dx\, d\nu \sum_f e^2_f\, q_f(x,\xi_A\,Q) \,
         \frac{d\sigma^{\gamma^* q}}{dx\,d\nu} \, D_f^h(z,\xi_A\,Q)  
         \, {\cal N}_A(z,\nu) \ .
\end{equation} 

We compare our theoretical predictions 
with the Hermes data.

\begin{figure}[tb]
\begin{center}
\parbox{9cm}{
\epsfig{figure=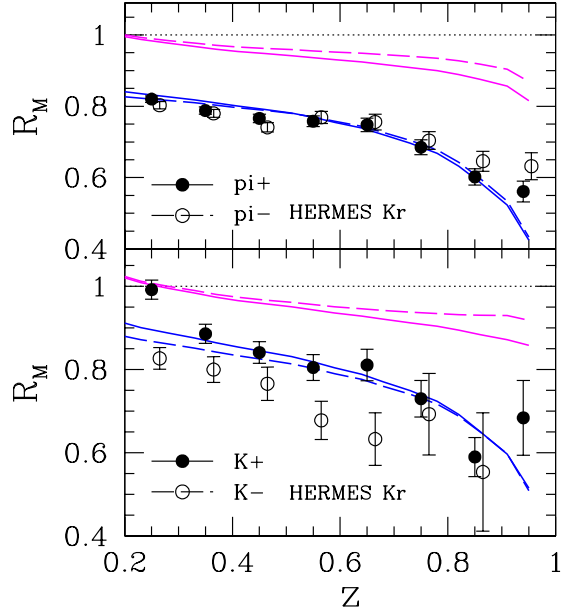,width=7.5cm
,clip=,bbllx=27pt,bblly=150pt,bburx=510pt,bbury=680pt}

}
\caption{\footnotesize
$z$-distributions for charge- and flavour-separated hadrons
at HERMES  on Krypton target. The upper pair of curves includes
rescaling without absorption, and the lower pair rescaling plus
Bialas-Chmaj absorption with an effective string tension $0.4$ GeV/fm.}
\label{fig:Hermes'02Kr}
\end{center}
\vskip-.4cm
\end{figure}

The rescaling plus absorption model
gives the respective multiplicity ratios shown in
ref. \cite{Hermes}. 
In fig. 2 one sees that the general agreement is
rather satisfactory, only
the $K^-$ spectra are not accurately predicted.
The very exact Hermes data demand a consideration of the details
of fragmentation. 
The $K^-$ mesons contain $\bar u s$ quarks which are  not
valence quarks of the proton, therefore they can only be created as 
second rank hadrons. We are currently considering differences in
kaon, pion, proton and antiproton spectra which are due to different
fragmentation processes.

In summary, the rescaling plus nuclear absorption model 
\cite{Accardi:2002tv} 
can describe  the HERMES data on the nuclear
modification of hadron production in DIS 
processes quite well. A more reasonable choice of parameters with the vacuum
string tension and a smaller prehadronic cross section does not
change this property. The A-dependence in the absorption model is
simple. If there is total absorption, then  the number of 
produced hadrons is proportional to $A^{2/3}$ for large nuclei since only
hadrons from the away surface of the nucleus can contribute
to the cross section. The nuclei analysed so far experimentally are
not
so heavy, then the A-dependence becomes z- and $\nu$ dependent. 
To assess the usefulness of this variable, quantitative analyses in
different models are needed. If one studies the $Q^2$
dependence of the multiplicity ratios one sees a weak rise of $R_M$ at small
$Q^2$ followed by an extended plateau. This points to gluon radiation 
before the prehadron is formed. 
Probably  in most cases of the Hermes experiment the prehadron is
produced
 rather
quickly and the gluon is radiated only along 
a short path, but the radiation 
has to be analysed in order to derive the needed prehadron
cross section  from color transparency.
Finally in high energy nucleus nucleus collisions many more challenges 
lie ahead. The hope is that one can
use the knowledge from deep inelastic scattering in nuclei 
to deduce carefully the properties of the
new matter which the particles have traversed.

\vspace{1cm}

{\bf Acknowledgments}
 
\vspace{0.15cm}

I am very grateful to A. Accardi, V. Muccifora and B.~Kopeliovich
for discussions. I thank all organizers, especially A. Bialas 
for the very stimulating meeting in Cracow. 
This work is partially funded by the exchange program between the
universities of Cracow and Heidelberg.

\end{document}